\documentclass[11pt]{article}
\usepackage[T1]{fontenc}
\usepackage[latin1]{inputenc}
\usepackage[dvips]{epsfig}
\usepackage{graphicx}
\usepackage[english]{babel}
\usepackage{amsmath}
\usepackage{amssymb}
\usepackage{amsfonts}
\usepackage{color}
\DeclareMathOperator{\arctanh}{arctanh}
\DeclareMathOperator{\arccoth}{arccoth}
\DeclareMathOperator{\arcsinh}{arcsinh}
\DeclareMathOperator{\arccosh}{arccosh}
 
\def\beq{\begin{equation}}
\def\eneq{\end{equation}}
\def\bea{\begin{eqnarray}}
\def\enea{\end{eqnarray}}

\begin{document}

\title {\large{ \bf   $(2+1)$-dimensional $f(R)$ gravity solutions
via Hojman symmetry }}
\author {  \small{ \bf F. Darabi}\hspace{-1mm}{ \footnote{ Corresponding author. e-mail: f.darabi@azaruniv.ac.ir}} , \small{ \bf  M. Golmohammadi}\hspace{-1mm}{
                \footnote{ author. e-mail: golmohammadi@azaruniv.ac.ir
        }}, \small{ \bf A. Rezaei-Aghdam }\hspace{-1mm}{\footnote{author. e-mail: rezaei-a@azaruniv.ac.ir}} \\
        {\small{\em Department of Physics, Azarbaijan Shahid Madani University}}\\
        {\small{\em   53714-161, Tabriz, Iran  }}}
\maketitle
\begin{abstract}
        In this paper, we use the Hojman symmetry approach 
        to find  new 
        $(2+1)$-dimensional $f(R)$ gravity solutions, in comparison to Noether
symmetry
approach. In the special case of Hojman symmetry vector $X=R$, we recover
 $(2+1)$-dimensional BTZ black hole and generalized
$(2+1)$-dimensional BTZ black hole solutions, obtained by Noether
symmetry
approach, and  the interesting point is that  the cosmological constant is appeared as the direct manifestation of Hojman symmetry. 
\end{abstract}

\maketitle
\section{Introduction}

 The  $(2+1)$-dimensional gravity has no black hole solutions for vanishing cosmological constant \cite{Ida}.  However, BTZ black hole solutions have been obtained for $(2+1)$-dimensional gravity with a negative cosmological constant, defined by the following action  \cite{BTZ} and \cite{BHTZ} 
\begin{equation}\label{ac1} 
I=\frac{1}{2}\int dx^{3} \sqrt{-g}\,
(R-2\Lambda),
\end{equation}
where $\Lambda=-l^{-2}$ is  characterized by a  length scale $l$. The line element in $(t, r, \phi)$ coordinates is given by\begin{equation}\label{metric}
ds^2 =-f(r)dt^2 + f^{-1}(r)dr^{2}+r^2\left(d\phi
-\frac{J}{2r^2}dt\right)^2, 
\end{equation} 
\begin{equation}\label{metric2}
f(r)=\left(-m+\frac{r^2}{l^2} +\frac{J^2}{4
r^2}\right)
\end{equation} 
where the mass $m$ and angular momentum $J$ are two parameters corresponding to time displacement and
rotational symmetries associated with two Killing vectors $\partial_t$ and $\partial_\phi$, respectively.
Unlike the asymptotically flat Schwarzschild and Kerr black hole solutions  having curvature singularity at $r=0$, the BTZ black hole is asymptotically anti-de-Sitter (AdS) having no curvature singularity at $r=0$. A BTZ black hole with $J\neq 0$ describes a spacetime with a constant negative
curvature having outer and inner horizons $r_{+}$ (event horizon) and $r_{-}$ (Cauchy horizon) respectively, given by  
\begin{equation}
r^{2}_{\pm}=\frac{l^2}{2}\left(m\pm\sqrt{m^2 -
\displaystyle{\frac{J^2}{l^2}} }\right). \label{horizon1} 
\end{equation} 
{It is known that at spacetime dimensions lower than 4, alternative
theories of gravity are needed to define a proper Newtonian limit \cite{Romero} and \cite{SPGT}. This necessity becomes remarkable when one intends to study the lower dimensional
black holes, like the well known BTZ black holes. Among the well-known generalization of General Relativity, like
Brans-Dicke  gravity \cite{BD} scalar-tensor gravity \cite{ST} and \cite{3}, {$f(T)$ gravity \cite{WGW} and \cite{AD}} and Lovelock gravity \cite{Lov},
the  $f(R)$ gravity \cite{JCAP}-\cite{CDe} is one of those alternative theories of gravity which is
under recent focus of attention. Using suitable conformal transformations, $f(R)$ gravity becomes equivalent to  a scalar-tensor gravity and this remarkable
feature makes it more interesting among the other alternatives of general relativity. Motivated by the above mentioned features of $f(R)$ gravity, a Noether symmetry approach has already been developed by the authors \cite{DAR}
to obtain $(2+1)$ dimensional black hole solutions in the framework of $f(R)$ gravity.}

An alternative  approach,  so called
Hojman symmetry approach, has recently been received attention by which one
can  find new exact solutions \cite{Hoj}-\cite{Myrzakul}. 
Unlike the Noether symmetry
approach which needs Lagrangian and Hamiltonian functions, in the Hojman
symmetry approach we just need the symmetry
vectors and the corresponding conserved charges
which are easily obtained by using the equations of motion.

{In the present paper, we intend to revisit the previous problem, considered
in \cite{DAR},} and obtain possible $(2+1)$ dimensional BTZ black hole solutions in the context of $f(R)$ gravity, using Hojman symmetry approach, as an alternative   to  the Noether symmetry approach. {This study is of particular importance
  because some new $(2+1)$ dimensional generalized BTZ black hole solutions  of $f(R)$ gravity, as well as corresponding symmetry vectors,  may be obtained which are absent in the Noether symmetry approach. This motivates 
us to apply Hojman symmetry approach for other gravitational systems and
 enables us to find new features which are not reported in their Noether symmetry approach.}
\section{$(2+1)$-dimensional $f(R)$ gravity } 

The action  for $(2+1)$-dimensional $f(R)$ gravity
is given by
\begin{equation}\label{1}
I=\frac{1}{2}\int d^3x \sqrt{-g}f({R}).
\end{equation}
 We consider the line element in the following form \cite{BTZ}
\begin{equation}\label{2}
ds^2=[-N^2(r)+r^2M^2(r)]dt^2+{N^{-2}(r)}{dr^2}
+2r^2M(r)dt d\phi+r^2d\phi^2,
\end{equation}
where the radial functions $N(r)$ and $M(r)$ are   considered as degrees
of freedom.  The  Ricci scalar is obtained
\begin{equation}\label{3}
R=-\frac{1}{2r}(4rN'^2+4rNN''-r^3M'^2+8NN'),
\end{equation}
where $'$ denotes the derivative with respect to $r$.
Using the method of Lagrange multipliers to set $R$ as a constraint of the dynamics,  generalizing
the degrees of freedom and defining a canonical  Lagrangian ${\cal L}={\cal L}(N, M, R, {N'}, {M'}, {R'})$, the action (\ref{1}) casts in the following form

\bea\label{4}
{\cal S}=\int d^3x \sqrt{-g}[f({R})-\lambda(R+\frac{1}{2r}(4rN'^2+4rNN''
-r^3M'^2+8NN'))].
\enea
After variation  with respect to $R$, we find $\lambda = f_R\equiv df/dR$
 and the action is rewritten as
\bea\label{5}
{\cal S}=\int d^3x \sqrt{-g}[f({R})-f_R(R+\frac{1}{2r}(4rN'^2+4rNN''
-r^3M'^2+8NN'))].
\enea
Integrating by parts lead to the   Lagrangian 
\begin{equation}\label{6}
{\cal L}=r(f-R f_R)+\frac{r^3}{2}f_R M'^2-2f_R NN'+2rf_{RR}R'NN',
\end{equation}
where $f_{RR}\equiv d^2f/dR^2$. The Euler-Lagrange equations  for $N$, $M$ and $R$ are derived respectively as
\begin{equation}\label{7}
N(f_{RRR}R'^2+f_{RR}R'')=0,
\end{equation}
\begin{equation}\label{8}
({r^3}f_R M')'=0,
\end{equation}
\bea\label{9}
-rRf_{RR}+\frac{r^3}{2}f_{RR}M'^2-4f_{RR}NN'-2rf_{RR}N'^2
-2rf_{RR}NN''=0.
\enea

\section{   Hojman symmetry approach}
Consider a set of second-order ordinary differential equations
\begin{eqnarray}\label{force}
\ddot{q}^i=F^{i}(q^j,\dot{q}^j,t),\qquad i,j=1,\,2,\,...\,n
\end{eqnarray}
where $q^{i}$ and $F^{i}$ denote the generalized coordinates and forces,  respectively, and each over dot denotes  derivative with respect to time $t$. If there exists an associated symmetry vector $X^i=X^i(q^j,\dot{q}^j,t)$, then it should satisfy the  differential equation
\cite{Hoj}
\begin{equation}\label{sym.vector}
\frac{\mathrm{d}^2X^i}{\mathrm{d}t^2} - \frac{\partial F^i}{\partial q^j}\, X^j - \frac{\partial F^i}{\partial \dot{q}^j} \, \frac{\mathrm{d}X^j}{\mathrm{d}t} =\,0,
\end{equation}\\
where\\
\begin{equation}\label{d/dt}
\frac{ \mathrm{d}}{\mathrm{d}t}\,=\,\frac{\partial}{\partial t}+\dot{q}^i \, \frac{\partial}{\partial q^j} \, +{F}^i \, \frac{\partial}{\partial \dot q^i}.
\end{equation}\\
The symmetry vector $X^i$ maps the solutions $q^i$ of Eq.(\ref{force}) into the solutions $\hat{q}^i$
of the same equations (up to $\epsilon^2$ terms), under the infinitesimal transformation
\begin{equation}\label{infi. trans}
\nonumber
\hat{q}^i=q^i+\epsilon X^i \left( q^j, \dot{q}^j, t \right).
\end{equation}
Using this property, the Hojman conserved quantities are defined by the following theorem \cite{Hoj} and \cite{Hoj1}:\\
\\ {\bf Theorem}:\\
1. Provided that   $F^i$ satisfies
  
 \begin{equation}\label{cons. quantity0}
\frac{\partial F^i}{\partial \dot{q}^i} =0 \, ,
\end{equation}
then

\begin{equation}
Q=\frac{\partial X^i}{\partial q^i}+ \frac{\partial }{\partial \dot{q}^i} \left( \frac{\mathrm{d}X^i}{\mathrm{d}t}\right),
\end{equation}
is a conserved quantity. \\
\\
2.  Provided that $F^i$ satisfies

\begin{equation}{\label{gamma}}
\frac{\partial F^i}{\partial \dot{q}^i} =-\frac{\mathrm{d}}{\mathrm{d}t}\ln \gamma,
\end{equation}
then
\begin{equation}\label{cons. quantity1}
Q =\frac{1}{\gamma} \frac{\partial (\gamma X^i)}{\partial q^i}+ \frac{\partial }{\partial \dot{q}^i} \left( \frac{\mathrm{d}X^i}{\mathrm{d}t}\right),
\end{equation}
is a conserved quantity, where $\gamma$ is merely a function of $q^i$.\\

\section{ (2+1)-dimensional $f(R)$ gravity solutions via Hojman Symmetry}
     The equations of motion that are obtained in  section 2, can be rewritten
in accordance with Hojman symmetry approach as follows 
\begin{equation}{\label{R''}}
R''=-h(R)R'^2,   
\end{equation}

\begin{equation}{\label{mr}}
        r^3f_R\,M'=C,
\end{equation}
$C$ being a constant, and
\begin{equation}{\label{nr}}
        -rRf_{RR}+\frac{r^3}{2}f_{RR}M'^2-4f_{RR}NN'-2\,rf_{RR}N'^2-2\,rf_{RR}NN''=0,
\end{equation}
where 
\begin{equation}{\label{hr}}
h(R)=\frac{f_{RRR}}{f_{RR}}.
\end{equation}
{Note that $h(R)$ should be well-defined for each proposed
$f(R)$}. 
By comparing (\ref{R''}) and  (\ref{force}), we can recognize that $F(R, R')=-h(R){R'}^2$. Also from Eq.(\ref{gamma}), we obtain

\begin{equation}
\gamma(R)=\gamma_0\,e^{\int2h(R)\mathrm{d}R},
\end{equation}
where $\gamma_0$ is a constant.
If there is no explicit dependence of $X$ on $r$, namely $X=X(R,R')$, then we can rewrite Eqs. (\ref{sym.vector}) and  (\ref{cons. quantity1}), respectively
as
\begin{equation}{\label{symmetryX}}
         \left( \frac{\partial\,^2X}{\partial R^2}+h_R X+h(R)\frac{\partial X}{\partial R}\right)+R'\,^2h^2(R)\frac{\partial\,^2X}{\partial R'\,^2}-R'\left(2h(R)\frac{\partial\,^2X}{\partial R\,\partial R'}+h_R\frac{\partial X}{\partial R'}\right)=0,
\end{equation} 
and 
\begin{equation}{\label{cons. quntity2}}
        Q =\frac{1}{\gamma} \frac{\partial (\gamma X)}{\partial R}+ \frac{\partial }{\partial R'} \left( \frac{\mathrm{d}X}{\mathrm{d}r}\right),
\end{equation}
where $h_R\equiv\frac{\mathrm{d}h}{\mathrm{d}R}$.
In general, solving the differential equation (\ref{symmetryX}) for vector $X$ is difficult. In this regard, we will limit ourselves to some particular ansatz for vector $X$, proposed in Ref. \cite{Hoj1},
as follows.

 \subsection{$X \sim X(R)$}
 From Eqs.(\ref{symmetryX}) and (\ref{cons. quntity2}), we see that 
 
 \begin{equation}{\label{Q/2}}
        h(R)X+\frac{\mathrm{d}X}{\mathrm{d}R}=\frac{Q}{2},
 \end{equation}
 where $Q$ is a conserved quantity. 
 
 \subsubsection{$X = R$}
 As a first step, if we simply consider $X=R{\neq 0}$ then from equation (\ref{Q/2}) {we find a well defined}  
 
 \begin{equation}
 h(R)=\frac{Q}{2R}-\frac{1}{R},
 \end{equation} 
 and from equations ({\ref{R''}}) and ({\ref{hr}}) we have
 \begin{equation}\label{30}
 f(R)=C_1+C_2R+C_3R^{\frac{Q}{2}+1},
 \end{equation}
 and
 \begin{equation}{\label{rr1}}
 R(r)=\frac{1}{2^{\frac{2}{Q}}}\left[ Q\left( C_5+C_4r\right) \right] ^{\frac{2}{Q}}.
 \end{equation}
  Finally, from equations  {(\ref{mr}), (\ref{30}) and ({\ref{rr1}})} for $M(r)$ and setting $C_5=0$ for simplicity, we obtain
\begin{equation}\label{mr1}
        M(r)=-\frac{CC_6^2C_4^2\ln(C_2+C_6C_4r)}{C_2^3}+\frac{CC_6^2C_4^2\ln(r)}{C_2^3}+\frac{CC_6C_4}{C_2^2r}-\frac{1}{2}\frac{C}{C_2r^2}+C_7,
 \end{equation}  
where $C_6=\frac{Q}{2}C_3(\frac{1}{2}Q+1)$. {Note that the requirement $R(r)\neq 0$ for a well-defined $h(R)$, excludes the origin $r= 0$ from the domain of $r$ for $Q>0$, and avoids $r\rightarrow \infty $  for $Q<0$. Therefore, at least there is no curvature singularity issue at the origin}. 

Using (\ref{30}), (\ref{rr1}) and (\ref{mr1}) in the equation of motion (\ref{nr}) we obtain
 \begin{equation}{\label{nr1}}
\begin{split}
        & N^2(r)=-\frac{1}{4C_2^4 r^2} \left[2C^2C_6C_4r \ln (-\frac{C_2}{C_6C_4r}-1)(3C_6C_4r+2{C_2})\right]+2C_8\frac{1}{r}+\frac{3}{2}\frac{C^2C_6C_4}{C_2^3}\frac{1}{r}\\
&\qquad~~~~~ 
{-\frac{Q^2}{2(Q+1)(3Q+2)}\left( \frac{C_4Q}{2}\right) ^{\frac{2}{Q}} r^{\frac{2}{Q}+2}}+\frac{C^2}{4C_2^2}\frac{1}{r^2}-2C_9
 \end{split}
 \end{equation} 
 where $C_8$ and $C_9$ are constants of integration. Now (\ref{mr1}) and (\ref{nr1}) determine the spherically symmetric solutions for the metric (\ref{2}) subject to the  Ricci scalar (\ref{rr1}).
These solutions are obtained by imposing Hojmam symmetry. Note that since $f(R)$ is not appeared in the field equations (\ref{mr}) and (\ref{nr}), the constant  $C_1$ is not appeared in the solutions for $M(r)$ and $N(r)$ as a direct consequence of imposing the Hojman symmetry along $X = R$.

{In order to compare
these solutions with those of obtained in \cite{DAR}, by imposing Noether symmetry,  we may consider the solution (51) in \cite{DAR} and set   {$C_3=0$} here to recover
 $f(R)=C_1+C_2R$ which is linear in terms of $R$ similar to $f(R)=R+D_3$, in  \cite{DAR}. This provides us with the following solutions
\begin{equation}\label{mr1'}
        M(r)=-\frac{1}{2}\frac{C}{C_2r^2}+C_7,
 \end{equation} 
\begin{equation}\label{34}
 N^2(r)=\frac{C^2}{4C_2^2}\frac{1}{r^2}+ 2\frac{C_8}{r}{-\frac{Q^2}{2(Q+1)(3Q+2)}\left( \frac{C_4Q}{2}\right) ^{\frac{2}{Q}} r^{\frac{2}{Q}+2}}-2C_9.       
 \end{equation}
{where $Q>0$ and $C_4>0$ are assumed.} A comparison between the solutions} (43)\footnote{We mean the solution (43)
 for
 which $D_1=0, D_2=1$ are imposed. }  and  (51) in  \cite{DAR} and the solutions (\ref{mr1'}) and (\ref{34})
here, shows that both solutions
are the same, up to some identifications between the constants in each solution. 

{In order to investigate the black hole property of these solutions,  we write
the metric (\ref{2})  in the following  form
\begin{equation}\label{47}
ds^2=-N^2(r)dt^2+N^{-2}(r)dr^2+r^2[M^2(r)dt+d\phi]^2.
\end{equation}
For the given constants in the shift function,  we may set  $N^2(r)=0$ 
to find the horizons of the black hole metric (\ref{47}). Therefore, the spherical solutions (\ref{mr1}) and (\ref{nr1}) are capable of being as a black hole solution for $f(R)$ gravity (\ref{30}), subject to the non-vanishing Ricci scalar
(\ref{rr1}), provided that their asymptotic behaviours are analyzed.
}

Now, we investigate on the possibility of recovering the well known BTZ black
hole solution
\begin{equation}\label{41''}
M(r)=-{\frac{J}{2r^2}},
\end{equation}
\begin{equation}\label{49'}
N^2(r)=-m+\frac{r^{2}}{l^{2}}+\frac{J^2}{4r^2},
\end{equation}
where $m$ and $J$ are the mass and angular momentum of the
black hole, respectively. In this regard, we may use the following initial identifications 
\begin{equation}
C=J,~~~C_1=2/l^2,~~~C_2=1,~~~C_7=C_8=0,~~~C_9=m/2.
\end{equation}
At this stage, it is left to identify the third term in (\ref{34}) with the
term ${r^{2}}/{l^{2}}$. To this end, we may consider $C_4=2Q^{-1}(\mp6Q^{-2})^{Q/2}$
and  
$Q\sim l$, assuming  $l\rightarrow \infty$,
which gives us the   third term in (\ref{34}) with the behavior ${\pm r^{2}}/{l^{2}}$.  Therefore, the solutions  (\ref{mr1'}) and (\ref{34}) exhibit an asymptotically de Sitter and anti-de Sitter spacetimes, corresponding
to ${- r^{2}}/{l^{2}}$ and ${+ r^{2}}/{l^{2}}$, respectively.

One may conclude that the regular BTZ black hole is a solution corresponding to  $f(R)=C_1+R$ equipped with a Hojman
symmetry vector $X=R$, {and the corresponding gauge freedom}  in choosing the constant term $C_1$ is fixed by a cosmological constant $\Lambda<0$ as $C_1={\pm}2l^{-2}{={\mp}2\Lambda>0}$. {Although $\Lambda$
appears as a gauge choice here, however,} similar to the constants $m$ and $J$ which are {interpreted as} the conserved charges corresponding to the time translation
and rotation symmetries, respectively,  we may {interpret} the  
{cosmological constant $\Lambda$} as the conserved charge {(because of $Q\sim (-\Lambda)^{-1/2}$)} corresponding to the {Hojman} symmetry of solutions under the infinitesimal displacement of $R$. This is
an important result accounting for the fact that the cosmological constant, {in the regular BTZ black hole}, is nothing but the manifestation of Hojman symmetry {which induces the symmetry vector $X=R$ and generates the conserved
charge $Q\sim (-\Lambda)^{-1/2}$}.  

By assuming $C_7=0, C_8 \neq 0$ we find \begin{equation}\label{41'''}
M(r)=-{\frac{J}{2r^2}},
\end{equation}
\begin{equation}\label{49''}
N^2(r)=-m+\frac{r^{2}}{l^{2}}+ 2\frac{C_8}{r}+\frac{J^2}{4r^2},
\end{equation}
which is the same generalized BTZ
black hole solution which was obtained in \cite{DAR} using Noether symmetry.

Now, we may analyze the asymptotic structure of the generalized BTZ solutions  (\ref{mr1'}) and (\ref{34}) in the limit $r\rightarrow \infty$
to check whether or not the solutions represent asymptotically anti-de Sitter
spacetime. It is easy to show that  by choosing $C_7=0$ and  non-vanishing constants $C, C_2, C_4, C_8$ and $C_9$, the solutions  (\ref{mr1'}) and (\ref{34}) exhibit an asymptotically de Sitter and anti-de Sitter spacetime, provided
that $C_4=2Q^{-1}(-6Q^{-2})^{Q/2}$
and $C_4=2Q^{-1}(+6Q^{-2})^{Q/2}$, respectively,
assuming $Q\sim l$, $l\rightarrow \infty$. Therefore, we may interpret the solutions
(\ref{mr1'}) and (\ref{34}) as an almost generalized BTZ black hole which
is asymptotically de Sitter and anti-de Sitter
spacetime with   asymptotic cosmological
constants $\Lambda\sim l^{-2}>0$ and $\Lambda\sim -l^{-2}<0$, respectively.

 \subsubsection{$X= \lambda \tan (R),~~~\lambda=$const}
 
 According to the previous subsection, by fixing $Q=2\lambda$,
we find  
{ 
\begin{equation}
        h(R)=-\frac{\sin R}{\cos R},
\end{equation}}
\begin{equation}\label{35}
        f(R)=C_1+C_2R+C_3{\cos R},
 \end{equation}
 and 
 \begin{equation}\label{36}
        R(r)=\arcsin(C_4r+C_5).
 \end{equation}
For simplicity, we fix $C_5=0$, {for which there is no curvature singularity at the origin
$r=0$. Again, the requirement $R(r)\neq (2k+1)\pi/2$ for a well-defined $h(R)$, excludes some definite values from the domain of $r$.} 

Then, we have
 \begin{equation}\label{37'}
        M(r)=-\frac{CC_3^2C_4^2}{C_2^3} \ln(-C_2+C_3C_4r)+\frac{CC_3^2C_4^2}{C_2^3}\ln{r}-\frac{C}{2C_2}\frac{1}{r^2}-\frac{CC_3C_4}{C_2^2}\frac{1}{r}+C_6.
 \end{equation}
Using (\ref{35}), (\ref{36}) and (\ref{37'}) in the equation of motion (\ref{nr}) we obtain
    \begin{equation}{\label{nr2}}
         \begin{split}
        & N^2(r)=-2C_7+\frac{C^2}{4C_2^2}\frac{1}{r^2}+2C_8\frac{1}{r}-\frac{3C^2C_3C_4}{2C_2^3}\frac{1}{r}-\sqrt{1-C_4^2r^2}\,\frac{1}{36C_4^3r}(5C_4^2r^2-8)\\
        &\qquad~~~      
        +\frac{1}{4C_2^4r^2}\left[2C^2C_4C_3r\ln(C_3-\frac{C_2}{C_4r})  (-3C_3C_4r+2C_2)\right]-\frac{1}{12C_4^2}(2C_4^2r^2-3)\arcsin(C_4r),
        \end{split}
        \end{equation}
        where $C_7$ and $C_8$ are integration constants.
For the given constants in the shift function,  we may set  $N^2(r)=0$ 
to find the horizons of the black hole metric (\ref{47}). Therefore, the spherical solutions (\ref{37'}) and (\ref{nr2}) are capable of being as a black hole solution for $f(R)$ gravity (\ref{35}) subject to the Ricci scalar
(\ref{36}) {satisfying the condition  $R(r)\neq (2k+1)\pi/2$}.

For the special case $C_3=0$, namely $f(R)=C_1+C_2R$, we obtain
\begin{equation}\label{370}
        M(r)=-\frac{C}{2C_2}\frac{1}{r^2}+C_6,
 \end{equation} 
 \begin{equation}\label{371}
        N^2(r)=\frac{1}{12C_4^2} (2C_4^2r^2-3)\arcsin(C_4r) -\frac{1}{36C_4^2}\frac{-8+5\,C_4^2r^2}{C_4r}\sqrt{1-C_4^2r^2}+\frac{C^2}{4C_2^2}\frac{1}{r^2}+2C_8\frac{1}{r}-2C_7,
 \end{equation}{where the domain of $r$ is limited by the inequality $1-C_4r\geq
 0.$ }
Unlike the previous solutions  (\ref{mr1'}) and (\ref{34}), the solutions  (\ref{370}) and (\ref{371}) are new in comparison to the
solutions (43) and (51) obtained for the same $f(R)=C_1+C_2R$ gravity  in \cite{DAR}.

Now, we investigate on the possibility of recovering the well known BTZ black
hole solution from the solutions  (\ref{370}) and (\ref{371}). It turns
out that the existence of first and second terms in (\ref{371}),
subject to $C_4\neq 0$, does not allow for such recovery even if we set   $C_6=C_8= 0, C_7=m/2$.
 Therefore, rather than BTZ black hole or generalized
BTZ black hole solutions, the Hojman symmetry provides us  with the   {new
 (2+1)
dimensional  $f(R)$ gravity  solution}, in comparison
        to the solution which has already been obtained by Noether symmetry in  \cite{DAR}.

{ The study of asymptotic structure shows that
by choosing $C_6=0,$ together with  finite-value constants $C, C_2, C_4, C_7$ and $C_8$, the solutions (\ref{370}) and (\ref{371}) cannot exhibit an asymptotically anti-de Sitter spacetime, because of the ill-defined trigonometric function, in the first
term, and also the imaginary behaviour of the square
root,  in the second term, for $C_4r>1$. 

However, by
choosing an infinitesimal $C_4$ and restricting the domain  $0\leq r\leq C_4^{-1}$,
it is possible to find the asymptotic behaviour up to the limit
$r\rightarrow C_4^{-1}$ ($C_4r\rightarrow1$) where the first and second terms goes
like  constant terms, while the third and forth terms almost vanishes. Therefore, provided that $C_4$ is infinitesimal  and the domain of $r$ is restricted by  $0\leq r\leq C_4^{-1}$, we may interpret the solutions
(\ref{370}) and (\ref{371}) as new
generalized (2+1)
dimensional  $f(R)$ gravity  solution
which is asymptotically flat spacetime.

  }

{
\subsubsection{$X= \lambda \tanh (R),~~~\lambda=$const}
 By this ansatz, again from the procedure mentioned in the previous subsections and setting $Q=2\lambda$, we have 
\begin{equation}
h(R)=\frac{\sinh R}{\cosh R} ,
\end{equation}
\begin{equation}\label{500}
f(R)=C_1+C_2R+C_3\cosh R,  \end{equation} 
and
\begin{equation}\label{510}
R(r)=\arcsinh{(C_4r+C_5).}
\end{equation}
 By fixing $C_5=0$, for simplicity,
we find that there is no curvature singularity at the origin
$r=0$.  Moreover, the function $h(R)$ is well defined for the whole range
of coordinate $r$. Finally, we obtain
\begin{equation}\label{520}
M(r)=-\frac{CC_3^2C_4^2}{C_2^3} \ln(C_2+C_3C_4r)+\frac{CC_3^2C_4^2}{C_2^3}\ln{r}-\frac{C}{2C_2}\frac{1}{r^2}+\frac{CC_3C_4}{C_2^2}\frac{1}{r}+C_6.
\end{equation} 
Using (\ref{500}), (\ref{510}) and (\ref{520}) in the equation of motion (\ref{nr}), we find
\begin{equation}{\label{nr30}}
\begin{split}
& N^2(r)=-\frac{1}{4C_2^4r^2}\left[ 2C^2C_3C_4r\ln \left( \frac{C_2}{C_4r}+C_3\right) \left( 2C_2+3C_3C_4r\right) \right]+\frac{8+5C_4^2r^2}{36C_4^3r}\sqrt{1+C_4^2r^2} \\
&\qquad~~~~~~
-\frac{3+2C_4^2r^2}{12C_4^2}\arcsinh(C_4r)+\frac{C^2}{4C_2^2}\frac{1}{r^2}+\frac{3C^2C_3C_4}{2C_2^3}\frac{1}{r}+2C_8\frac{1}{r}-2C_7,
\end{split}
\end{equation}
where $C_7$ and $C_8$ are integration constants. The horizons of  black hole metric (\ref{47})
are obtained by setting $N^2(r)=0$, for the given constants in the shift function. Therefore, the spherical solutions (\ref{520}) and (\ref{nr30}) are capable of being as a black hole solution for $f(R)$ gravity (\ref{500}) subject to the Ricci scalar
(\ref{510}) with $C_5=0$.
{For the special case $C_3=0$,  $f(R)=C_1+C_2R$, we obtain 
        \begin{equation}\label{M(r)}
        M(r)=-\frac{C}{2C_2}\frac{1}{r^2}+C_6,
        \end{equation}
        \begin{equation}\label{NNr0}
        N^2(r)=\frac{8+5C_4^2r^2}{36C_4^3r}\sqrt{1+C_4^2r^2}-\frac{3+2C_4^2r^2}{12C_4^2}\arcsinh(C_4r)+\frac{C^2}{4C_2^2}\frac{1}{r^2}+2C_8\frac{1}{r}-2C_7.
        \end{equation}

These solutions are also new for $f(R)=C_1+C_2R$ gravity, because         of the first two new terms of $N^2(r)$, in comparison to the solutions  obtained by Noether symmetry in  \cite{DAR}. However, similar to the previous
        cases,  the existence of first and second terms in (\ref{NNr0}),
        subject to $C_4> 0$, does not allow for  recovering the BTZ black hole, even if we set   $C_6=C_8= 0$.
        Therefore, rather than BTZ black hole or generalized
BTZ black hole solutions, the Hojman symmetry provides us  with the   {new
 (2+1)
dimensional  $f(R)$ gravity  solution}, in comparison
        to the solution which has already been obtained by Noether symmetry in  \cite{DAR}.}}

The study of asymptotic structure at $r \rightarrow \infty$
shows that the third, forth and fifth terms of $N^2(r)$ can be ignored, the first term  
can behave as $-\Lambda r^2\equiv r^2/l^2$, and the second term  represents an uncommon behaviour. Therefore this solution can partially exhibit an asymptotically anti-de Sitter spacetime, due to the first term. 

 However,  the presence of    second term including  $``\arcsinh(C_4r)$'' may represent some other features of this solution which deserves attention.
Actually, if  a de Sitter term  ${-\Lambda r^2\equiv r^2/l^2}$
 dominates the other terms at $r\rightarrow \infty$, it will give rise to an asymptotic
anti-de Sitter spacetime with a  cosmological constant having any given value
$\Lambda\sim -l^{-2}$, from  $\Lambda_{as}\rightarrow 0^-$ ($l\rightarrow \infty$) to  $\Lambda_{as}\rightarrow -\infty$  ($l\rightarrow 0$).  Similarly, if  a de Sitter term  $\Lambda r^2\equiv r^2/l^2$
 dominates the other terms at $r\rightarrow \infty$, it will give rise to an asymptotic
de Sitter spacetime with a  cosmological constant having any given value
$\Lambda\sim l^{-2}$, from  $\Lambda_{as}\rightarrow 0^+$ ($l\rightarrow \infty$) to  $\Lambda_{as}\rightarrow \infty$  ($l\rightarrow 0$). 
 So, one may think that the second term at $r\rightarrow \infty$ can be effectively considered as ``$-\Lambda(r)r^2$'' where
$\Lambda(r)\equiv\arcsinh(C_4r)$  plays the role of a positive
cosmological
constant with an asymptotic  value $\Lambda_{as}\rightarrow \infty$ at $r\rightarrow \infty$, or as ``$\Lambda(r)r^2$'' where
$\Lambda(r)\equiv-\arcsinh(C_4r)$  plays the role of a negative
cosmological
constant with an  asymptotic  value $\Lambda_{as}\rightarrow -\infty$ at $r\rightarrow \infty$. Therefore, the
solutions \eqref{M(r)}  and \eqref{NNr0} exhibit an asymptotically de Sitter
and anti-de Sitter spacetime with an effective asymptotic values of cosmological
constant $\Lambda+\Lambda_{as}>0$ and $\Lambda+\Lambda_{as}<0$, respectively.
{\subsubsection{$X= \lambda \coth (R),~~~\lambda=$const}
        By this ansatz, and setting $Q=2\lambda$, we have 
        \begin{equation}
        h(R)=\frac{\cosh (R)}{\sinh (R)},
        \end{equation}
\begin{equation}\label{5000}
f(R)=C_1+C_2R+C_3\sinh R,  
\end{equation} 
and
\begin{equation}\label{5100}
R(r)=\arccosh {(C_4r+C_5).}
\end{equation}
By taking $C_5=0$, 
the function $h(R)$ is not well-defined at $C_4r=1$, hence the
domain of $r$ is limited to $r>1/C_4$ and  the problem of curvature singularity at the origin
$r=0$ is completely removed.  Then, we have
\begin{equation}\label{5200}
M(r)=-\frac{CC_3^2C_4^2}{C_2^3} \ln(C_2+C_3C_4r)+\frac{CC_3^2C_4^2}{C_2^3}\ln{r}-\frac{C}{2C_2}\frac{1}{r^2}+\frac{CC_3C_4}{C_2^2}\frac{1}{r}+C_6.
\end{equation} 
Using (\ref{5000}), (\ref{5100}) and (\ref{5200}) in the equation of motion (\ref{nr}), we obtain
\begin{equation}{\label{nr33}}
\begin{split}
& N^2(r)=-\frac{1}{4C_2^4r^2}\left[ 2C^2C_3C_4r\ln \left( \frac{C_2}{C_4r}+C_3\right) \left( 2C_2+3C_3C_4r\right) \right]+\frac{-8+5C_4^2r^2}{36C_4^3r}\sqrt{-1+C_4^2r^2} \\
&\qquad~~~~~~
+{\frac {\ln  \left( C_{{4}}r+
                \sqrt {{C_{{4}}}^{2}{r}^{2}-1} \right) }{4{C_{{4}}}^{2}}}
 -\frac{r^2}{6}\arccosh(C_4r)+\frac{C^2}{4C_2^2}\frac{1}{r^2} +\frac{3C^2C_3C_4}{2C_2^3}\frac{1}{r}+2C_8\frac{1}{r}-2C_7,\\
 \end{split}
\end{equation}}
{where $C_7$ and $C_8$ are integration constants.
For the given constants in the shift function,  we may set  $N^2(r)=0$ 
to find the horizons of the black hole metric (\ref{47}). Therefore, the spherical solutions (\ref{5200}) and (\ref{nr33}) are capable of being as a black hole solution for $f(R)$ gravity (\ref{5000}) subject to the Ricci scalar
(\ref{5100}) with $C_5=0$.
For the special case $C_3=0$, namely $f(R)=C_1+C_2R$, we obtain 
        \begin{equation}\label{Mr61}
        M(r)=-\frac{C}{2C_2}\frac{1}{r^2}+C_6,
        \end{equation}
        \begin{equation}\label{NNr-1}
        \begin{split}
        &       N^2(r)=\frac{-8+5C_4^2r^2}{36C_4^3r}\sqrt{C_4^2r^2-1}-\frac{r^2}{6}\arccosh(C_4r)+{\frac {\ln  \left( C_{{4}}r+
                        \sqrt {{C_{{4}}}^{2}{r}^{2}-1} \right) }{4{C_{4}}^{2}}}\\
                &\qquad~~~~~~
        +\frac{C^2}{4C_2^2}\frac{1}{r^2}+2C_8\frac{1}{r}-2C_7,
        \end{split}
        \end{equation}
where the domain of $r$ is limited by the inequality $C_4r\geq 1.$        These solutions are also new for $f(R)=C_1+C_2R$ gravity, because of the first and second terms of $N^2(r)$, in comparison to the solutions  obtained by Noether symmetry in  \cite{DAR}. Similar to the previous
        cases,  the existence of first and second terms in (\ref{NNr-1}),
        subject to $C_4\neq 0$, does not allow for  recovering the BTZ black hole, even if we set   $C_6=C_8= 0$.
       Therefore, rather than BTZ black hole or generalized
BTZ black hole solutions, the Hojman symmetry provides us  with the   {new
 (2+1)
dimensional  $f(R)$ gravity  solution}, in comparison
        to the solution which has already been obtained by Noether symmetry in  \cite{DAR}. }

The study of asymptotic structure at $r \rightarrow \infty$
shows that the first term of $N^2(r)$ 
behaves as $-\Lambda r^2\equiv r^2/l^2$, but the second and third terms  represent  uncommon behaviours. Therefore this solution can partially exhibit an asymptotically anti-de Sitter spacetime, due to the first term, though the presence of  the second and third terms with   uncommon asymptotic behaviours may represent some other features of this solution. 

The presence of   negatively signed second term 
with  $``\arccosh(C_4r)$''   may be considered at $r\rightarrow \infty$ as ``$-\Lambda(r)r^2$'' where
$\Lambda(r)\equiv\arccosh(C_4r)$ plays the role of an asymptotic positive
cosmological
constant with a value $\Lambda_{as}\rightarrow \infty$. The presence of   positively signed third term 
including ``$\ln  \left( C_{{4}}r \right)$'' may be considered at $r\rightarrow \infty$ as ``$-\Lambda'(r)r^2$'' where
$\Lambda'(r)\equiv {-\ln  \left(2 C_{{4}}r \right) }/{r^{2}}$ plays the role of an effective asymptotic negative
cosmological
constant with a value $\Lambda'_{as}\rightarrow 0^{-}$.  Therefore, the
solutions \eqref{Mr61}  and \eqref{NNr-1} exhibit an asymptotically de Sitter spacetime with an effective asymptotic  value of cosmological
constant $\Lambda+\Lambda_{as}+\Lambda'_{as}>0$.

\subsubsection{$X= \lambda \sinh (R),~~~\lambda=$const}

 By this ansatz, again from the procedure mentioned at the previous subsections and setting $Q=2\lambda$, we have 
{\begin{equation}
                h(R)=-\frac{\cosh (R)-1}{\sinh (R)},
 \end{equation}} 
 \begin{equation}\label{42}
 f(R)=C_1+C_2R+C_3\left[ -\ln {(\tanh(\frac{1}{2}R)-1)}-\ln {(\tanh(\frac{1}{2}R)+1)}\right],  \end{equation} 
and
\begin{equation}\label{43}
        R(r)=-2\arctanh {(C_4r+C_5).}
\end{equation}
Finally, by fixing $C_5=0$ for simplicity,
we obtain
\begin{equation}\label{44}
        M(r)=-\frac{CC_3^2C_4^2}{C_2^3} \ln(-C_2+C_3C_4r)+\frac{CC_3^2C_4^2}{C_2^3}\ln{r}-\frac{C}{2C_2}\frac{1}{r^2}-\frac{CC_3C_4}{C_2^2}\frac{1}{r}+C_6.
\end{equation} 
{Note that by choosing $C_5=0$, the requirement $R(r)\neq 0$ for a well-defined $h(R)$, excludes $r= 0$ and  the problem of curvature singularity at the origin
is completely removed.} Using (\ref{42}), (\ref{43}) and (\ref{44}) in the equation of motion (\ref{nr}), we obtain
\begin{equation}{\label{nr0}}
        \begin{split}
& {N^2(r)={-\frac{1}{4C_2^4r^2}\left[ 2C^2C_3C_4r\ln \left( -\frac{C_2}{C_4r}+C_3\right) \left( -2C_2+3C_3C_4r\right) \right]}-\frac{1}{3C_4^3}\ln \left( C_4^2r^2-1\right)\frac{1}{r}+}  \\
&\qquad~~~~~
{+\frac{r^2}{3} {\arctanh}(C_4r)-\frac {1}{C_4^2}\arccoth(C_4r)+\frac{2}{3C_4}r+\frac{C^2}{{4}C_2^2}\frac{1}{r^2}-\frac{3C^2C_3C_4}{{2C_2^3}}\frac{1}{r}+2C_8\frac{1}{r}-2C_7,}
        \end{split}
        \end{equation}        
where $C_7$ and $C_8$ are integration constants.
For the given constants in the shift function,  we may set  $N^2(r)=0$ 
to find the horizons of the black hole metric (\ref{47}). Therefore, the spherical solutions (\ref{44}) and (\ref{nr0}) are capable of being as a black hole solution for $f(R)$ gravity (\ref{42}) subject to the Ricci scalar
(\ref{43}) with  $C_5=0$.
{For the special case $C_3=0$, namely $f(R)=C_1+C_2R$, we obtain 
\begin{equation}\label{Mr68}
M(r)=-\frac{C}{2C_2}\frac{1}{r^2}+C_6,
\end{equation}
\begin{equation}\label{NNr69}
        {N^2(r)={-}\frac{1}{3C_4^3}\ln{(C_4^2r^2-1)}\frac{1}{r}{-\frac{1}{C_4^2}}\arccoth(C_4r)+\frac{r^2}{3}\arctanh(C_4r)+\frac{2}{3C_4}r+\frac{C^2}{4C_2^2}\frac{1}{r^2}+2C_8\frac{1}{r}-2C_7.}
\end{equation}
These solutions are not  physical  because the domains of $``\ln$'', $``\arccoth$''
 are not compatible with the domain of  $``\arctanh$'', in the solution $N^2(r)$.  

\subsection{$X = X(R')$}

For the choice $X=X(R')$,  the equation of symmetry  vector $X$, (\ref{symmetryX}), reads as Euler equation
\begin{equation}\label{Equation1}
h_RX+R'\,^2h^2(R)\frac{\mathrm{d}\,^2X}{\mathrm{d} R'\,^2}-R'h_R\frac{\mathrm{d} X}{\mathrm{d} R'}=0,
\end{equation}
from which $X$ and $h(R)$ are obtained respectively as
follows
\begin{equation}
X=A_1R'+A_2R'^n,
\end{equation}

\begin{equation}
h(R)=-\frac{1}{nR+h_0},
\end{equation}
where $A_1, A_2, h_0$ and $n$ are constant parameters, and the Hojman conserved quantity reads as

\begin{equation}{\label{conservedq}}
Q=2h(R) R'^n-h(R)n(n+1)R'^n.
\end{equation}
By using Eqs. (\ref{hr}) and (\ref{conservedq}) we have
\begin{equation}\label{51}
        f(R)=C_1+C_2\left( R+\frac{h_0}{n}\right) +C_3\left( R+\frac{h_0}{n}\right) ^{\frac{2n-1}{n}}
\end{equation}
and
\begin{equation}\label{52}
        R(r)=\frac{Q_0^{\frac{1}{n-1}}}{n}\left[ (1-n)(-r+C_4)\right] ^{\frac{n}{n-1}} -\frac{h_0}{n},
\end{equation}
which $Q_0=-\frac{Q}{2-n(n+1)}$. {
Note that by choosing $C_4=0$, the requirement $R(r)\neq -h_0/n$ for a well-defined $h(R)$, excludes $r= 0$ and  the problem of curvature singularity at the origin
is completely removed.
} 

From equations (\ref{mr}) and (\ref{nr}) we have

\begin{equation}\label{55}
 M(r)=-\frac{CC_3^2C_5^2}{C_2^3}\ln (C_2+C_3C_5r)+\frac{CC_3^2C_5^2}{C_2^3}\ln(r)-\frac{C}{2C_2}\frac{1}{r^2}+\frac{CC_3C_5}{C_2^2}\frac{1}{r}+C_6, \end{equation} 
where $C_5=\frac{2n-1}{n^{\frac{2n-1}{n}}}Q_0^{\frac{1}{n}}(n-1)$. {Moreover,  we may discard $n=1,-2$
cases which result in vanishing conserved charge $Q$.
}
Again, using equations (\ref{51}), (\ref{52}) and (\ref{nr}) we obtain
\begin{equation}\label{nr55}
\begin{split}
&
        N^2(r)= \frac{{C^2}}{4C_2^2}\frac{1}{r^2}+(2C_7+\frac{C^2C_3C_5}{2C_2^3})\frac{1}{r}-\frac{3C_3^2C_5^2C^2}{2C_2^4}-2C_8-\frac{Q_0^{\frac{1}{n-1}}}{n(3n-2)(4n-3)} \left( (n-1)r\right) ^{\frac{3n-2}{n-1}}\\
        &\qquad~~~~~
      {-\frac{1}{4C_2^4r^2}\left[2C^2C_3C_5r\ln \left( \frac{C_2}{(n-1)r}+\frac{C_3C_5}{n-1}\right) \left(3C_3C_5r+2C_2 \right)\right]  } .
\end{split}
\end{equation}
For the given constants in the shift function,  we may set  $N^2(r)=0$ 
to find the horizons of the black hole metric (\ref{47}). Therefore, the spherical solutions (\ref{55}) and (\ref{nr55}) are capable of being as a black hole solution for $f(R)$ gravity (\ref{51}) subject to the Ricci scalar
(\ref{52}).

 For $C_3=0$, namely $f(R)=C_1+C_2(R+\frac{h_0}{n})$, we obtain
\begin{equation}\label{mrr}
M(r)=-\frac{C}{2C_2}\frac{1}{r^2}+C_6,
\end{equation}
\begin{equation}\label{nnr}
N^2(r)={\frac{C^2}{4C_2^2}\frac{1}{r^2}-\frac{Q_0^{\frac{1}{n-1}}}{n(3n-2)(4n-3)} \left( (n-1)r\right) ^{\frac{3n-2}{n-1}}+2C_7\frac{1}{r}-2C_8}.   
\end{equation}  

{These solutions are not capable of being either BTZ black hole or generalized
BTZ black hole, with asymptotically anti-de Sitter spacetime, because
the power $\frac{3n-2}{n-1}$ in the second term cannot be equal to 2, unless
for $n=0$ which itself diverges the second term.
} Therefore, rather than BTZ black hole or generalized
BTZ black hole solutions, the Hojman symmetry provides us  with the   {new
 (2+1)
dimensional  $f(R)$ gravity  solution}, in comparison
        to the solution which has already been obtained by Noether symmetry in  \cite{DAR}.

The asymptotic behaviour of this solution is determined by the second term
including $r^{\frac{3n-2}{n-1}}$ which depends on the value of $n\neq 0$
as follows:
\begin{itemize}
\item For $n=2/3$ a constant term arises which contributes to the constant
mass term $-2C_8$,  

\item For $n=3/4$ a  term ``$r^{-1}$'' arises which contributes to the third term,
\item For $n=4/5$ a  term ``$r^{-2}$'' arises which contributes to the first term.
\end{itemize}  
\textcolor{red}{  }
For $n=1/2$ a novel linear ``$r$'' term arises which may describe    the flatness of the galaxy rotation curves \cite{Philip} and \cite{GHS} as well as the non-trivial contribution of the
characteristic feature of the surrounding quintessence field \cite{HD1}-\cite{HD3}.            
\subsection{$X(R,R')\sim R' g(R)$}
We consider another ansatz $X(R,R')\sim R' g(R)$, where $g(R)$ is an arbitrary function. By this assumption, in order for $X$ to be the symmetry vector,  we obtain
 \begin{equation}{\label{39}}
        h(R)=\frac{g_{RR}}{g_R}
 \end{equation}
 Also, the Hojman conserved quantity is obtained as
 \begin{equation}\label{Q_0}
 Q_0=R'g_R.
 \end{equation}
Now, by considering some ansatzs for $g(R)$, we find some exact solutions in the following.
\\

\subsubsection{ $g(R)=\lambda\, e^{\alpha R}$}          

By this choice and using equation (\ref{hr}), (\ref{39}) and (\ref{Q_0}), we obtain
{\begin{equation}
                h(R)=\alpha,
\end{equation}}
\begin{equation}\label{61}
        f(R)=C_1+C_2R+C_3e^{\alpha R},
\end{equation}
and
\begin{equation}\label{62}
        R(r)=\frac{1}{\alpha}\ln \left( \frac{Q_0(r+C_4)}{\lambda}\right),
\end{equation}
{where $\alpha$ is constant.}
By fixing $C_4=0$, for simplicity, it turns out that there is no curvature
singularity at the origin $r=0$.
We have
\begin{equation}\label{63}
        M(r)=-\frac{CC_3^2\alpha ^2Q_0^2}{\lambda ^2C_2^3} \ln ({\frac{C_2\lambda}{r}+C_3\alpha Q_0})-\frac{1}{2}\frac{C}{C_2}\frac{1}{r^2}+\frac{CC_3\alpha Q_0}{\lambda C_2^2}\frac{1}{r}+C_5,
\end{equation}
and 
\begin{equation}\label{64}
        \begin{split}
        &
        N^2(r)={-\frac{1}{4C_2^4\lambda^2r^2}\left[ 2\alpha Q_0C^2C_3r\ln \left( \frac{C_2\lambda}{Q_0r}+C_3\alpha\right) \left( 2C_2\lambda+3C_3\alpha Q_0r\right) \right]}\\
&\qquad~~~~~   
 {-\frac{r^2}{6\alpha}\ln \left( \frac{Q_0r}{\lambda}\right) }+\frac{5}{36\alpha}r^2+\frac{3Q_0C^2C_3\alpha}{2C_2^3\lambda}{\frac{1}{r}}+\frac{C^2}{4C_2^2}\frac{1}{r^2}+2C_7\frac{1}{r}-2C_6.
        \end{split}
\end{equation}
For the given constants in the shift function,  we may set  $N^2(r)=0$ 
to find the horizons of the black hole metric (\ref{47}). Therefore, the spherical solutions (\ref{63}) and (\ref{64}) are capable of being as a black hole solution for $f(R)$ gravity (\ref{61}) subject to the Ricci scalar
(\ref{62}).

For $C_3=0$, namely namely $f(R)=C_1+C_2R$, we obtain
\begin{equation}\label{Mr287}
        M(r)=-\frac{1}{2}\frac{C}{C_2}\frac{1}{r^2}+C_5,
\end{equation}
\begin{equation}\label{NNr288}
       N^2(r)=-\frac{r^2}{6\alpha}\ln (\frac{Q_0r}{\lambda})+\frac{C^2}{4C_2^2}\frac{1}{r^2}+2{C_7}\frac{1}{r}+\frac{5}{36\alpha}r^2-2{C_6}.
\end{equation}

The above solutions are  new, because of the extra first term $r^2\ln (\frac{Q_0r}{\lambda})$, in comparison to the solutions  obtained by Noether symmetry in  \cite{DAR}, for  $f(R)=C_1+C_2R$ gravity. Similar to the previous
cases,  the existence of $``\ln$'' term in (\ref{NNr288}) does not allow for recovering the  BTZ black hole, even if we set   $C_5= C_7=0$.
 Therefore, rather than BTZ black hole or generalized
BTZ black hole solutions, the Hojman symmetry provides us  with the   {new
 (2+1)
dimensional  $f(R)$ gravity  solution}, in comparison
        to the solution which has already been obtained by Noether symmetry in  \cite{DAR}.

The study of asymptotic structure at $r \rightarrow \infty$
shows that the forth term of $N^2(r)$ 
behaves as $-\Lambda r^2\equiv r^2/l^2$, 
but the first term    represents an  uncommon asymptotic behaviour. Therefore, the existence of the forth term (assuming $\alpha>o$) can partially exhibit an asymptotically anti-de Sitter spacetime.

The presence of   first term with $\alpha>0$ may be considered at $r\rightarrow \infty$ as ``$-\Lambda(r)r^2$'' where
$\Lambda(r)\equiv (1/6\alpha){\ln  \left(Q_0r/\lambda \right) }$ plays the role of an effective asymptotic positive
cosmological
constant with a behaviour $\Lambda_{as}\rightarrow \infty$.  Therefore, the
solutions \eqref{Mr287}  and \eqref{NNr288} exhibit an asymptotically de Sitter spacetime with an effective value of asymptotic  cosmological
constant $\Lambda+\Lambda_{as}>0$.

\subsubsection{$g(R)=\frac{(g_0+R)^{1+\alpha}}{1+\alpha}$}
Here $g_0$ and $\alpha$ are constants parameters. Again, from Eqs. (\ref{39}), (\ref{Q_0}), (\ref{mr}) and (\ref{nr}) we find respectively  $f(R)$, $R(r)$, $M(r)$ and $N^2(r)$ for the case $C_4=0$, as follows
{\begin{equation}
                h(R)=\frac{\alpha}{g_0+R},
\end{equation}}
\begin{equation}{\label{fr6}}
        f(R)=C_1+C_2(g_0+R)+C_3(g_0+R)^{\alpha+2},
\end{equation}

\begin{equation}{\label{rr6}}
R(r)=\left[ Q_0(1+\alpha)r\right] ^{\frac{1}{1+\alpha}}-g_0.
\end{equation}
{Note that  the requirement $R(r)\neq -g_0$ for a well-defined $h(R)$ excludes $r= 0$, and  the problem of curvature singularity at the origin
is completely removed.} We have

\begin{equation}{\label{mr6}}
M(r)=-\frac{CC_3^2C_5^2}{C_2^3} \ln ({\frac{C_2}{r}+C_3C_5})-\frac{1}{2}\frac{C}{C_2}\frac{1}{r^2}+\frac{CC_3C_5}{ C_2^2}\frac{1}{r}+{C_6},
\end{equation}
and 
 \begin{equation}\label{nr7}
        \begin{split}
        &
        N^2(r)={-\frac{1}{4C_2^4r^2}\left[ 2C^2C_3C_5r\ln \left( \frac{C_2}{Q_0(1+\alpha)r}+\frac{C_3C_5}{Q_0(1+\alpha)}\right)\left( 2C_2+3C_3C_5r\right) \right]}\\
        &\qquad~~~
         {-\frac{Q_0^{\frac{1}{1+\alpha}}}{(2\alpha+3)(3\alpha+4)}\left[  (1+\alpha)r\right]  ^{\frac{3+2\alpha}{1+\alpha}}}+\frac{g_0}{6}r^2+\frac{C^2}{{4}C_2^2}\frac{1}{r^2}+\left( 2{C_8}+\frac{C^2C_3C_5}{2C_2^3}\right) \frac{1}{r}\\
         &\qquad~~~
         -2{C_7}-\frac{3C^2C_3^2C_5^2}{2C_2^4},
        \end{split}
        \end{equation}       
        where $C_5=(\alpha+1)(\alpha+2)Q_0$.
For the given constants in the shift function,  we may set  $N^2(r)=0$ 
to find the horizons of the black hole metric (\ref{47}). Therefore, the spherical solutions (\ref{mr6}) and (\ref{nr7}) are capable of being as a black hole solution for $f(R)$ gravity (\ref{fr6}) subject to the Ricci scalar
(\ref{rr6}).

For the case $ C_3=0$, namely namely $f(R)=C_1+C_2R$, we obtain
\begin{equation}\label{mr6'}
M(r)=-\frac{1}{2}\frac{C}{C_2}\frac{1}{r^2}+{C_6},
\end{equation}
\begin{equation}\label{nnr6'}
N^2(r)=-\frac{Q_0^{\frac{1}{1+\alpha}}\left[ (\alpha +1)r\right] ^{\frac{2\alpha +3}{1+\alpha}}}{(2\alpha +3)(3\alpha +4)}+\frac{C^2}{4C_2^2}\frac{1}{r^2}+2C_8\frac{1}{r}+\frac{g_0}{6}r^2-2C_7.
\end{equation}

{These solutions are  considered as new, because of the first term,  in comparison to the solutions  obtained by Noether symmetry in  \cite{DAR} for  $f(R)=C_1+C_2R$ gravity.} These solutions are not capable of being either BTZ black hole or generalized
BTZ black hole because
the power $\frac{2\alpha +3}{1+\alpha}$ in the first term cannot be equal to 2.
Therefore, rather than BTZ black hole or generalized
BTZ black hole solutions, the Hojman symmetry provides us  with the   {new
 (2+1)
dimensional  $f(R)$ gravity  solution}, in comparison
        to the solution which has already been obtained by Noether symmetry in  \cite{DAR}.

The study of asymptotic structure at $r \rightarrow \infty$
shows that the existence of the fourth
$r^2$ term  in $N^2(r)$ can partially exhibit  asymptotically de Sitter
and anti-de Sitter spacetimes according to $g_0<o$ and $g_0>o$,
respectively. The first term of $N^2(r)$ 
  represents an $r$ dependent  behaviour   which is determined by the chosen value of $\alpha$, some examples of which are given as follows:

\begin{itemize}
\item For $\alpha=-3/2$ a constant term arises which contributes to the constant
mass term $-2C_7$,  

\item For $\alpha=-4/3$ a  term ``$r^{-1}$'' arises which contributes to the third term,
\item For $\alpha=-5/3$ a  term ``$r^{-2}$'' arises which contributes to the second term.
\end{itemize}

These three  cases are in agreement with the asymptotic behaviour exhibiting an asymptotically de Sitter or anti-de Sitter spacetime.
Note that no value of $\alpha$ can give rise to a term ``$r^2$'' to contribute
to the fourth
$r^2$ term. 

For $\alpha=-2$ a novel linear ``$r$'' term arises which may describe    the flatness of the galaxy rotation curves \cite{Philip} and \cite{GHS} as well as the non-trivial contribution of the
characteristic feature of the surrounding quintessence field \cite{HD1}-\cite{HD3}.

\subsubsection{$g(R)=\lambda \ln R$}
For this ansatz we have
{\begin{equation}
                h(R)=-\frac{1}{R},
\end{equation}}
\begin{equation}\label{73}
        f(R)=C_1+C_2R+C_3R\ln R,
\end{equation}

\begin{equation}\label{74}
R(r)=C_4 e^{\frac{Q_0r}{\lambda}}.
\end{equation}
{ The requirement $R(r)\neq 0$ for a well-defined $h(R)$, does not impose any restriction on the domain of $r$. Also, there is  no curvature singularity  at the origin $r=0$.} We have

\begin{equation}\label{75}
M(r)=-\frac{CC_3^2Q_0^2}{\lambda ^2C_5^3}\ln (\frac{C_5\lambda}{r}{+C_3Q_0})-\frac{C}{2C_5}\frac{1}{r^2}+\frac{CC_3Q_0}{\lambda C_5^2}\frac{1}{r}+C_6,
\end{equation}
where $C_5=C_2+C_3(1+\ln C_4)$, and
\begin{equation}\label{76}
        \begin{split}
        &
        N^2(r)={-\frac{1}{4C_5^4\lambda^2r^2}\left[ 2Q_0C^2C_3r\ln \left( \frac{C_5\lambda}{Q_0r}+C_3\right) \left( 3C_3Q_0r+2C_5\lambda\right) \right] }-\frac{C_4\lambda^2e^{\frac{Q_0r}{\lambda}}}{Q_0^3}\left( {Q_0-\frac{2\lambda}{r}}\right)\\
        &\qquad~~~~~~
        {+}\frac{C^2}{4C_5^2}\frac{1}{r^2}+\frac{3Q_0C^2C_3}{2C_5^3}\frac{1}{r}+2C_7\frac{1}{r}-2C_8.
        \end{split}
        \end{equation}
For the given constants in the shift function,  we may set  $N^2(r)=0$ 
to find the horizons of the black hole metric (\ref{47}). Therefore, the spherical solutions (\ref{75}) and (\ref{76}) are capable of being as a black hole solution for $f(R)$ gravity (\ref{73}) subject to the Ricci scalar
(\ref{74}).

For the case $ C_3=0$, namely namely $f(R)=C_1+C_2R$, we obtain
\begin{equation}\label{Mr3101}
M(r)=-\frac{C}{2C_5}\frac{1}{r^2}+C_6,
\end{equation}
\begin{equation}\label{NNr3102}
N^2(r)=-\frac{C_4\lambda^2}{Q_0^3}\frac{(-2\lambda +Q_0r)e^{\frac{Q_0}{\lambda}r}}{r}+\frac{C^2}{4C_5^2}\frac{1}{r^2}+2{C_7}\frac{1}{r}-2{C_8}.
\end{equation}

{These solutions are  considered as new, because of the first term,  in comparison to the solutions  obtained by Noether symmetry in  \cite{DAR} for  $f(R)=C_1+C_2R$ gravity. These solutions are not capable of being either BTZ black hole or generalized
BTZ black hole, due to the absence of $r^2$ term.
Therefore,  the Hojman symmetry provides us with the new  (2+1)
dimensional  $f(R)$ gravity  solution. 

The first term of $N^2(r)$ 
  represents an $r$ dependent  behaviour   and is determined by the chosen values of $C_4, \lambda, Q_0 $ which may lead this term to be positive or
negative.
A positive term 
 may be considered at $r\rightarrow \infty$ as ``$-\Lambda'(r)r^2$'' where
$\Lambda'(r)\equiv {-e^{\frac{Q_0}{\lambda}r}}/{r^{2}}$ plays the role of an effective asymptotic negative cosmological
constant with a value $\Lambda'_{as}\rightarrow -\infty$. A negative term 
 may be considered at $r\rightarrow \infty$ as ``$-\Lambda'(r)r^2$'' where
$\Lambda'(r)\equiv {e^{\frac{Q_0}{\lambda}r}}/{r^{2}}$ plays the role of an effective asymptotic positive cosmological
constant with a value $\Lambda'_{as}\rightarrow \infty$.   Therefore, the
solutions \eqref{Mr3101}  and \eqref{NNr3102} may exhibit  asymptotically de
Sitter and anti-de Sitter spacetimes,  with an effective values of asymptotic  cosmological
constant $\Lambda'_{as}>0$ and $\Lambda'_{as}<0$, respectively.

\section{Conclusions}

We have obtained  $(2+1)$-dimensional spherically
symmetric solutions in the context
 of  $(2+1)$-dimensional $f(R)$ gravity by using the Hojman symmetry. These
solutions are new in comparison to those obtained by Noether symmetry approach
\cite{DAR}.
 In the special case of Hojman symmetry along $X = R$,  these solutions cast in the form of $(2+1)$ dimensional BTZ black hole and  generalized $(2+1)$ dimensional BTZ black holes which have already been obtained by Noether symmetry approach
\cite{DAR}. The interesting point of Hojman symmetry approach is that  the cosmological constant is appeared as the direct manifestation of Hojman symmetry. {We aim to study the T- Duality problem \cite{eghbali}
of the obtained $(2+1)$ dimensional BTZ black hole solutions, 
in the next future.}

\end{document}